\documentstyle[12pt]{article}

\begin{document}

\author{ A. Lakshminarayan and N.L. Balazs \\{\sl Department of Physics,
S.U.N.Y. at Stony Brook,} \\ {\sl Stony Brook, N.Y. 11794.}}
\date{}
\title{On the Quantum Cat and Sawtooth Maps- Return to Generic Behaviour}
\maketitle
\newcommand{\newc}{\newcommand}
\newc{\beq}{\begin{equation}}
\newc{\eeq}{\end{equation}}
\newc{\beqa}{\begin{eqnarray}}
\newc{\eeqa}{\end{eqnarray}}
\newc{\longra}{\longrightarrow}
\newc{\longla}{\longleftarrow}
\newc{\ul}{\underline}\newc{\rarrow}{\rightarrow}
\newc{\br}{\langle}
\newc{\kt}{\rangle}
\newc{\hs}{\hspace}

\begin{abstract}
The quantization of the continous cat maps on the torus has led to
rather pathological quantum objects [6]. The non-generic behaviour of this
model has led some to conclude that the Correspondence Principle fails
in this case [2]. In this note we introduce the quantum sawtooth models,
since this is the natural  family to which the cat maps belong. Thus, a simple
propagator depending on a parameter is constructed which for integer values
of the parameter becomes pathological quantum cat maps, while away from
integer values we find a return to the generic behaviour of non-integrable
quantum systems.
\end{abstract}

\section{Introduction}

	\hs{.2in} In this note we will quantize the sawtooth maps
[1].  These are generalizations of the cat maps [6].  The quantum cat maps
have many non-generic features, resulting from the periodicity of the
propagator. There have been speculations about the ``failure of the
correspondence principle'', using the quantum cat map and arguments
about their algorithmic complexity (or lack of it) [2].  The arguments
of Joseph Ford and his coworkers relies on the fact that the quantum
dynamics of the cat map is identical to the classical motion on a
rational grid. The periodicity of every rational grid in the case of
cat maps is a peculiarity, and results in the periodicity of the
quantum propagator. All the eigenvalues of the quantum cat map are
roots of unity. The eigenvalues are highly degenerate, and the work of
J.P. Keating [3] has shown that in the classical limit the spectrum
becomes infinitely degenerate, although very slowly.

Given the controversy generated by the quantum cat maps, it is natural
to ask what happens in the case of their most natural generalizations,
the quantum sawtooth maps. Surprisingly, to the best of our knowledge,
this has not been done, although the quantization itself is
straightforward.  The classical mechanics of the sawtooth maps has
been investigated by several authors [1,4], I.C. Percival in [5], with
a view towards studying transport in the presence of partial barriers
such as cantori. I.C. Percival and his coworkers have developed
symbolic dynamics for these maps [1], and N. Bird and F.  Vivaldi [4]
have found their periodic orbits. {\em We find that the quantized sawtooth
maps are not periodic, their eigenangles are all irrational multiples
of $2 \pi$, and they display level repulsion characteristic of quantum
chaotic systems}. We will scale the eigenangles by $2 \pi$, so that
when we talk of ``rational eigenangles'', it means that the
eiganangles are rational fractions of $2 \pi$.

We notice that there is a $\log $ time, a time when quantum
interference effects dominate and destroy the picture of wavepackets
propagating as classical Liouville phase space densities.  When $K$,
the real parameter in the family, is very close to an integer, the
operator is nearly periodic initially, but with increasing time, the
periodicity is lost.  Related to this is the fact that the slightest
perturbation of the cat map, (making $K$  not an integer), seems to
produce irrational eigenangles (when the sawtooth map is
hyperbolic). These form bands that are clustered around the
eigenangles of the nearby quantum cat map.  The band structure
disappears rapidly when we move away from the cat map.  The
differences between the quantum cat maps and sawtooth maps is allied
to the mathematical problem of the differences between the complete
and incomplete Gauss sums of number theory.

\section{The Classical Map}

	\hs{.2in} We will very briefly describe the classical map.
	Consider a free particle that is subjected to time periodic impulses
	due to a force $F(q)$, given by:
\beq
F(q)\,=\, K\,\mbox{Saw}(q),
\eeq
where
\beq
	\left. \begin{array}{lcl}
		\mbox{Saw}(q)& =& q \;\;\;\;( -1/2 \, \leq  \, q \, < \, 1/2),\\
		\mbox{Saw}(q)&=& \mbox{Saw}(q+1).
	\end{array}
	\right.
\eeq
 This gives the impulse the shape of a sawtooth, and the map its name
 [1]. The Hamiltonian for this system can be taken to be
\beq
H(q,p,t)\,=\, \frac{p^{2}}{2} \,-  \, \frac{K\,(\mbox{Saw}(q))^{2}}{2}
 \sum_{n=- \infty}^{\infty} \delta(t-n).
\eeq
The potential is periodic with period 1.
The Hamiltonian equations of motion give us the map
\beq
	\left. \begin{array}{ccl}
	q^{\prime} & = & q \,+ \,  p^{\prime}\\
	 p^{\prime} & = & p \, + \, K \, \mbox{Saw}(q).
	\end{array}
	  \right.
\eeq

When $K$ is not an integer this map has a discontinuity at
 half integer points. The sawtooth on
the torus is obtained by imposing periodic boundary conditions in both
$q$ and $p$. This means that we take the above map {\em mod}~1.  We have
followed I.C. Percival [1] and taken the phase space to be the
``chosen torus'', centered at the origin, rather than the usual torus.
Then there is only one discontinuity at the point $q = 1/2$. When $K$ is
an integer this discontinuity vanishes, as it gets ``dissolved'' by the
modulo operation; these are cat maps.

The important point we note is that the potential is already
periodicised. Cat maps can also be obtained from the {\em non
periodic} potential $ -K q^{2}/2$, but the sawtooth maps {\em cannot}.
To see this imagine that the infinite phase plane is tessellated
by  fundamental squares. Then for integer values of $K$ the linear map
\beq
	\left. \begin{array}{ccl}
	q^{\prime} & = & q \,+ \,  p^{\prime}\\
	 p^{\prime} & = & p \, + \, K \, q.
	\end{array}
	  \right.
\eeq
obtained from the unperiodicised potential $-K \, q^{2}/2$ takes
equivalent points to equivalent points. Two phase points are
equivalent if they differ by an integer vector. If we {\em retain}
this potential and proceed with the quantization of the Hamiltonian of
eqn.(3), requirements of periodicity will naturally force us to
restrict ourselves to cat maps. Indeed this is the procedure of Joseph
Ford and his coworkers [2], for although their quantization method is
general enough the chosen potential was restrictive.

The sawtooth map on the torus is unstable for $K \, > \, 0$ and $K \,
< \, -4$. The stable regime is a curious map filled with many elliptic
islands, this is illustrated in fig.1. K=0 is the case of a free
rotor. All the periodic points of the unstable maps are hyperbolic
[1]. Periodic orbits of cat maps can be used to find them [4].

\section{ The Quantum Sawtooth}

\subsection{ The Propagator}
 \hs{.2in} The quantization of the Hamiltonian, eqn. (3), after
imposing periodic boundary conditions on position and momentum, give
us the quantum sawtooth propagator. The quantization of J. Ford et.
al. [2] is itself our starting point.  We impose {\em periodic }
boundary conditions on the states. The Planck's constant $ \hbar$ is
 related to the dimensionality, $N$, of the finite Hilbert
space by the relation $ 2 \pi \hbar \,=\, N^{-1}$. The periodicised
position eigenstates are denoted as $| q_{n} \kt$ and the periodicised
momentum states are denoted as $|p_{m} \kt $, $m,
\,n\, = -N/2, \ldots , N/2-1$. The transformation functions are
discretised plane waves,
\beq
\br q_{n}| p_{m} \kt \,=\, \frac{1}{\sqrt{N}} e^{2 \pi i mn/N}.
\eeq
The position and momentum eigenvalues are $n/N\,$;$\, n=-N/2,\ldots, N/2-1$.

The unitary propagator obtained by integrating the Hamiltonian
of eqn.(3),  quantized canonically, over one time step is
\beq
	\hat{U}\, =\, \exp \left( -i \hat{p}^{2}/2\hbar \right) \, \exp \left(
i K      (\mbox{Saw}(\hat{q}))^{2} /2 \hbar\right).
\eeq
The first term of the R.H.S. of the above equation is the propagator
corresponding to the free rotation, we denote it as $\hat{U}_{0}$, the
second part arises from the ``kick'' or the impulse and is denoted as
$\hat{U}_{1}$.  $\hat{U}$ is still the propagator for the map on the
whole plane.  The restriction to a torus is achieved quantally by
requiring that the action of the unitary operator maintains the
periodicity of the discrete toral states ( that are Dirac delta
combs).  To implement this, first consider the action of the free
propagator $\hat{U}_{0}$ on the discrete toral states,

Thus consider,
\beq
\br q_{n}|\hat{U}_{0}|p_{m} \kt \, =\, e^{-i \pi m^{2}/N}.\, e^{2 \pi i mn/N}.
\eeq
Here we have used the relation $ 2 \pi \hbar \,=\, N$.  The
requirement that the above be periodic in both $m$ and $n$ with a
period $N$, implies that $N$ be even. We will henceforth assume that
this is the case. If we had chosen anti-periodic boundary conditions on
the states we would have required $N$ to be odd. Similarly the mixed
representation of the kick operator is
\beq
\br p_{m}|\hat{U}_{1}|q_{n} \kt \,=\, e^{i \pi N K (\mbox{Saw}(n/N))^{2}}
	. \, e^{-2 \pi i mn/N}.
\eeq
The periodicity in $n$ and $m$ follows immediately from the
periodicity of the sawtooth potential. We in particular do {\em not} require
that $K$ be an integer. J. Ford et. al. used identical quantization
procedures [2],  but as noted earlier, the potential was taken to be
(essentially) $-K q^{2}/2$, which leads to the factor $e^{i \pi K
n^{2}/N} $ as the first term in the R.H.S. of the above equantion.
Imposing periodicity on this factor would then lead to the restriction
of $K$ to the integers. In fact, as we have seen this is true even
classically, if we start with the harmonic oscillator potential,
instead of the nonlinear periodic sawtooth potential.

Such quantizations can be carried out for any periodic potential.
When the potential is a cosine, the map is the famous standard map.
The quantum propagator is an $N \times  N$ matrix, when restricted to
act on the Hilbert space of $N$ states. Then we have $\mbox{Saw}(n/N)
\,=\, n/N$.
We can now put together the operators $\hat{U}_{0}$ and $\hat{U}_{1}$,
and write the quantum sawtooth map in the position representation as
\beq
\br q_{n}|U| q_{n^{\prime}} \kt \,=\, \frac{1}{N} e^{i \pi K n^{\prime \, 2}/N
} . \sum_{k=-N/2}^{N/2-1} e^{2 \pi k(n-n^{\prime})/N} \, e^{-i \pi k^{2}/N}.
\eeq
 The sum above simplifies upon using the Poisson summation formula,
and we get the  final form of the propagator as
\beq
\br q_{n}|U| q_{n^{\prime}} \kt \,=\, \frac{e^{-i \pi/4}}{\sqrt{N}}
e^{i \pi K n^{\prime 2}/N}\, e^{i \pi (n-n^{\prime})^{2}/N}.
\eeq

We have dropped the hats for the operators on the torus, which are
simply finite unitary matrices. When $K$ is an integer the above is a
quantum cat map, otherwise it is a discontinuos quantum sawtooth map. The
propagator is thus a very simple one, and is the natural
generalization of the quantum cat maps of Hannay, Berry and Ford
[6,2]. The operator $U$ has all the features we have noted
in the introduction. For integer $K$ it is a periodic operator with
all rational eigenangles, i.e., there is an integer $n(N)$ ($n$ is not
to be confused with the position labelling) such that $U^{n(N)}\,=\,
e^{i \phi (N)} I_{N}$. Here $I_{N}$ is the $N \times N $ identity
matrix, and $\phi(N)$ is a real phase.  When $K$ is not an integer
and the operator is not periodic, all the eigenangles are irrational.

We have to qualify the last statement, since : a) it is a numerical
observation,  b) there are cases for $ -4\, < \, K \, < \, 0 $, when
there are some rational eigenangles. These correspond to stable
sawtooth maps, when $\cos ^{-1}(K+2/2)$ is a rational multiple of
$\pi$.  If there is an elliptic fixed point
at the origin for a linear map on the plane, then the eigenangles
form a harmonic oscillator spectrum, with eigenangles given by the
equation
\beq
       (l+1/2) \cos ^{-1}(tr/2),\;\;\; l=\ldots, -2,-1,0,1,\ldots.
\eeq
where $tr$ is the trace of the classical matrix describing the map (
for instance, see ref.7).  In the case of the sawtooth map the trace
is $K+2$.  If the stable fixed point at the origin has a large
elliptic island, see fig.1, which does not ``feel'' the nonlinearity
of the map, then many sequences of eigenangles of the operator $U$ are
well predicted by the above equation.

The classical map has the symmetry of reflection about the center of
the square $ (q \rarrow -q,\, p \rarrow -p)$. This symmetry is present
in the quantum operator $U$, as $U_{n n^{\prime}} \, =\, U_{-n
-n^{\prime}}$.  The choice of origin at the center of the square makes
the symmetry matrix have the form
\beq
 	P_{N} \, =\, \left( \begin{array}{cc}
		1 & 0\\
		0  & R_{N-1}
		\end{array}
		\right),
\eeq
	where $R_{N-1}$ is the $ N-1 \times N-1 $ matrix with 1 along
the secondary diagonal, and 0 elsewhere. Thus $(R_{N-1})_{mn} \, =
\,1$ if $m \, + \, n \, + \, 1\, =\, N-1$ and zero otherwise. Since
$P_{N}^{2}\,=\, I_{N}$, and $[ P_{N}, U]\,=\, 0$, the eigenstates of
the propagator can be separated according to their parity.  Any
eigenvector is of the form
\beq
	|\psi_{ \pm} \kt \, =\, \left( \begin{array}{c}
				\alpha \\
				| \psi_{1} \kt \\
			   \pm	R_{N/2-1}| \psi_{1} \kt
				\end{array}
				\right),
\eeq
		where $\alpha$ is the component $\br -N/2 | \psi_{ \pm} \kt $.
This implies that odd parity eigenstates should have $ \alpha \, =\, 0$.
Thus there are $N/2-1$ odd parity states and $N/2 +1$ even parity states.
When finding the nearest neighbour distribution of the eigenangles, we
will select only the eigenangles corresponding to odd parity states.

\subsection{Numerical Results}

		\hs{.2in} The most commonly used statistics is the that
of the nearest neighbour spacing. Shown in fig.2 are the nearest
neighbour spacing distribution for $ K\,=\, 2.25, 2.50, 2.55, 3.25$.
The level repulsion is apparent. The statistics is for the 149 odd
parity states, when $N=300$. Better statistics would require more
eigenangles, but the essential feature of level repulsion is clear
enough. The statistics when $K$ is very close to an integer value,
when the sawtooth map  is almost a cat map, is bound to be rather peculiar.
In these cases there is, as has been noted above, a band like
structure, so that levels cluster around well separated eigenangles.
When we move away from the integer value the band like structure
quickly gives way to a more uniformly spread distribution exhibiting
level repulsion. This is the case of the fig.3, where we have the
nearest neighbour spacing distribution for the $K = .01, .1, 3.01,
3.1$.

Classically there is a significant difference between the case when
$K$ is close to zero, and when $K$ is close to some other integer. In
the former case we have just moved away from the integrable free rotor
at $K=0$. The KAM theorem conditions are not met, as the sawtooth map
is not smooth, hence we have no deformed tori.  The map becomes
immediately globally chaotic for a positive $K$ value. However for
small positive $K$ there are significant partial barriers to
transport, cantori made of parts of the stable and unstable
manifolds. These cantori are less important as barriers when the $K$
value is large. The Poisson distribution of the eigenangles for
$K=0.01$ and the level repulsion for $K=3.01$, shown in fig.3 is a
quantum manifestation of this difference. Also compare the cases,
$K=.1$ and $K=3.1$. That the eigenvalue statistics can be affected by
classical transport properties has been exhibited before [8].  The
sawtooth maps provide another example for this phenomenon, which needs
more study.

 	Figs.4,5,6 show contours of the autocorrelation functions.  We
use the coherent states developed by M. Saraceno [9] adapting it to
periodic boundary conditions.  It is a phase space representation that
allows the classical structures of quantized toral maps to be more
easily identified.  If $|p,q \kt $ is such a coherent state, $p$ and
$q$ take values on the classical $N \times N$ rational grid of the
torus. It is a state that is highly concentrated at (p,q), in the
sense of a minimum uncertainty wavepacket. Thus the autocorrelation
$|\br p,q|U^{t}|p,q \kt |^{2}$ is the probablility that a wavepacket
initially centered at $(p,q)$ ``comes back'' to $(p,q)$ after $t$ time
steps. For quantum cat maps the autocorrelation is periodic in time,
due to the periodicity of the propagator itself.  However sawtooth
maps that are far from cat maps, show autocorrelations like other
quantum systems, such as the quantum baker's map [10].  Fig.4 shows
autocorrelations for the cat $K=2$.  fig.5 is the case of the sawtooth
$K=2.25$. fig.6 is the case of the sawtooth $K=.5$. The value of the
inverse of Planck's constant in all the figures is $N=48$.

In all the figures the map has been shifted to the usual torus $[0,1)
\times [0,1)$. This does not affect the quantum or classical system in
any essential way. The fixed point at the origin now gets shifted to
the point $(1/2,0)$. The quantum cat map of fig.4 behaves as expected,
it is periodic. This results in the ``emptiness'' of fig.4, n=8, when
the propagator fixes all wavepackets.  The periodicity of the
propagator is the {\em classical} periodicity of the $N \times N $
rational grid, consisting of partitions of equal area
[6,2]. F.J. Dyson and H. Falk [12] have given bounds of this
periodicity for the case of Arnold's cat map, corresponding to the
case of $K=1$. If $m_{N}$ is the period of the lattice (and of the
quantum propagator) a lower bound has been established as $m_{N} > [
\log N/
\log \lambda] $, (for more complete statement of bounds see ref.12). This
surprisingly coincides with the so called $\log $ time [11] when
classical-quantum correspondence breaks down due to interference.

Thus if we assume that such lower bounds are valid in other cat maps
(this is not so hard to believe as these bounds are derived from the
divisibilty properties of Fibonnacci numbers) we see that the
periodicity of the propagator must be a post log time phenomenon. This
is to be expected as the reconstruction of the wavefunctions under a
quantum cat map proceeds due to some kind of coherent interference.
In fig.4 the first two time steps show significant peaks at classical
periodic orbits and nowhere else. The log time for this case of $N=48,
K=2$ is 2. Hence there are strong interference effects after this
time, yet the structures produced are very regular, resulting finally
in the identity operator at $n=8$, apart from a phase factor.

 The post log time structures of the quantum cat map are thus very
special. There are too many classical periodic orbits such that if
each is assinged a phase space volume $h\; (= 1/N)$ there would be too
many to fit the unit square. Yet there are coherent structures around
the periodic orbits. For instance in fig.4, n$=4$, the uniform
striations are precicely the lines along which the classical orbits of
period 4 lie.  It is said that there ``is no $ \log (1/\hbar)$ problem
for the cat maps'' [3].  If we view the $\log $ time as heralding the
onset of quantum interference effects, we can only surmise that such
effects are very coherent and special for the cat map.

 For the sawtooth maps the $\log $ time surfaces in a generic
manner. The larger eigenvalue (inverse of the smaller one, as map
preserves area)  of the classical map is
\beq
	\lambda \,=\,\frac{ (2+K)+ \sqrt{K(K+4)}}{2}.
\eeq
The Lyapunov exponent, $\Lambda$ is $\log \lambda$. Thus when $K=2.25$
and $.5$ the exponents are $\Lambda_{2.25} = \log 4$ and $\Lambda_{.5}
= \log 2$.  The $\log $ time is then $ \log (48/2)/ \Lambda$. The
$\log $ time for $K=.5$ $\approx 5$ is twice that of the case
$K=2.25$. Thus we see that after the second time step (n=2), in fig.5
interference effects set in and the peaks that are at classical
periodic points are no longer clearly visible.  While the
corresponding situation for $K=.5$ in fig.6 shows the longer $\log $
time, and after about 5 time steps the interference effects are
visible. When $K$ is very close to an integer the correlations are
once more very close to that of the nearby cat map; with increasing
time, however, interference effects destroy the periodicity. Figs. 5
and 6 also show some coherent reconstructions after the $\log $ time,
but more study is needed to understand these.

 \section{Conclusions}

	In this note we have begun the study of quantum sawtooth
maps.  They place the non-generic quantum cat maps in a family with
generic behaviour. We have found level repulsion and the existence of
a $\log $ time in the sawtooth maps. The $\log $ times agree well with the
expectations.  The propagator becomes periodic when the sawtooth maps
become cat maps, otherwise they have no exact periodicity. Thus we
cannot expect that the periodic orbit sums for the sawtooth maps will be
exact. Unlike the cat maps such sums will once more be semiclassical
approximations.

	The singular character of the quantum cat maps is reflected
in the degeneracy of the eigenangles. While quantized chaotic systems
in  general do not have any degeneracies, the cat maps become
increasingly  degenerate  as we approach the classical limit.
The sawtooth maps have no such degeneracies and the classical limit
of this map may be expected to behave in the  usual manner.

\newpage

\begin{center}
 {\large{ \bf References}}
\end{center}

\begin{enumerate}

\item  I.C. Percival and F. Vivaldi, {\sl Physica D } {\bf 27}, 373 (1987).

\item  J. Ford, G. Mantica, G.H. Ristow, {\sl Physica D} {\bf 50}, 493 (1991).

\item  J.P. Keating, {\sl Nonlinearity} {\bf 4}, 301, 335 (1991).

\item  N. Bird and F. Vivaldi, {\sl Physica D} {\bf 30}, 164 (1988).

\item  M.J. Giannoni, A. Voros and J. Zinn-Justin, eds.,
{\sl Chaos and Quantum Physics, Proceedings of Les Houches Summer School},
 Session LII 1989, Elsevier, Amsterdam, (1990).

\item J.H. Hannay and M.V. Berry, {\sl Physica D} {\bf 1}, 267 (1980).

\item A. Lakshminaryan, Ph.D. Thesis, S.U.N.Y. at Stony Brook,
 May, 1993.

\item  F.M. Izrailev, {\sl Phys. Rep.} {\bf 196}, 299 (1990).

\item  M. Saraceno, {\sl Ann. Phys., (N.Y.)} {\bf 199}, 37 (1990).

\item  N.L. Balazs and A. Voros, {\sl Ann. Phys., (N.Y.)}{\bf 190}, 1
(1989).

\item  M.V. Berry and N.L. Balazs, {\sl J.Phys. A} {\bf 12}, 625 (1979).

\item F.J. Dyson and H. Falk, {\sl The Am. Math. Monthly} {\bf 99}, 603
(1992).
\end{enumerate}

\end{document}